\documentclass[mathleft, fleqn, final
]{an}
\usepackage{graphicx}
\usepackage{amssymb,amsmath}
\usepackage{times}
\begin{document}

\Pagespan{789}{}
\Yearpublication{2006}%
\Yearsubmission{2005}%
\Month{11}%
\Volume{999}%
\Issue{88}%

\title{Magnetic field decay in normal radio pulsars}

\author{ A.P. Igoshev\inst{1}
\and
S.B. Popov\inst{2}\fnmsep\thanks{Corresponding author:
  \email{sergepolar@gmail.com}\newline}
}
\titlerunning{Hall attractor and magnetic field decay in pulsars}
\authorrunning{A.P. Igoshev \& S. B. Popov}
\institute{
Department of Astrophysics/IMAPP
Radboud University Nijmegen
P.O. Box 9010
6500 GL Nijmegen
The Netherlands
\and 
Lomonosov Moscow State University, Sternberg Astronomical Institute,
Universitetski pr. 13,
119991 Moscow, Russia}

\received{30 May 2005}
\accepted{11 Nov 2005}
\publonline{later}


\abstract{%
We analyse the origin of the magnetic field decay in normal radio pulsars found by us
in a recent study. This decay has a typical time scale $\sim 4 \times
10^5$~yrs, and operates in the range $\sim 10^5$~--~few$\times 10^5$~yrs. We
demonstrate that this field evolution may be either due to the Ohmic decay
related to the scattering from phonons, or due to the Hall cascade which
reaches the Hall attractor. According to our analysis the first possibility
seems to be more reliable. So, we attribute the discovered field decay
mainly to
the Ohmic decay on phonons which is saturated at the age few$\times
10^5$~yrs, when a NS cools down to the critical temperature below which
the phonon scattering does not contribute much to the resistivity of the
crust. Some role of the Hall effect and attractor is not excluded, and will
be analysed in our further studies.}

\maketitle

\section{Introduction}
Observational appearances of most (if not all) types
of neutron stars (NSs) significantly depend on their magnetic fields.
The magnetic field properties include not only
its strengths, but also its topology.  
Evolution of both was a subject of
numerous studies 
(see an early review in \cite{chan1992}, 
and more recent in \cite{geppert2009}).

Theoretical models typically predict that the field might decay due to
several reasons.  However, analysis of observational data mostly yields
controversial results.  Especially, if we restrict our consideration to the
normal radio pulsars only. On one hand, decay of magnetar fields
seems to be inevitable for explanation of observational manifestations of 
these sources (see a review in Mereghetti, Pons \& Melatos 2015\nocite{mpm2such}). 
Another strong evidence in favour of field decay comes from the fact
that fields of millisecond (recycled) pulsars 
are much weaker than those of normal ones, which can be explained
by significant decay due to accretion in  low-mass X-ray binaries
(LMXBs).  On other hand, there is no definitive answer if standard
magnetic fields ($\sim10^{12}$ G) 
significantly decay on the time scale few million years 
-- the lifetime scale of a normal radio pulsar.

Recently we proposed a modified data analysis to probe the magnetic field decay in
normal radio pulsars (\cite{ip2014}). 
Application of the new method
resulted in discovery of magnetic field decay by a factor $\sim 2$ in the
time interval $\sim 10^5$~--~$3\times 10^5$~yrs. It was speculated that on longer
time scales this decay might stall in order to be in correspondence with previous
studies which had not found any significant field decay on longer time scales. 
In this paper we present preliminary results of our study on 
the reason for such non-uniform field evolution. The full consideration of the
problem will be presented elsewhere (\cite{ip2015})

\section{Field decay in radio pulsars}
In case of normal radio pulsars their magnetic field can not be directly measured.
However, the basic pulsar properties such as period and period derivative 
should already contain some information
about magnetic field evolution. Still, there is no straightforward 
way to extract the field evolution because both the kinematic magnetic field 
estimate $\beta B = \sqrt{P\dot P}$, where $\beta$ is a (probably time dependent)
coefficient determined by NS properties, 
and the spin-down age $\tau=P/(2\dot P)$ are related to each other. 
It means that there is no good independent age estimate applicable to significant
number of pulsars of different ages. 
Thus, it is necessary to use specific (statistical) 
methods to uncover magnetic field
evolution from the data based on model-independent observable parameters, such as
$P$ and $\dot P$.

In the previous study (\cite{ip2014}) we presented a new path towards 
breaking the dependency between the spin-down age and the field estimate.
A new method to study field evolution in normal
radio pulsars was proposed.
The method is extensively tested  with population synthesis
technique using synthetic populations with different magnetic field
evolution.

Our method is similar to the well-known pulsar current approach
(\cite{vn1981}), 
but we study the ``flow'' of pulsars not along the direction of growing spin
periods, but along the spin-down age.
We propose to count a number of pulsars $N_i$
in spin-down age intervals $\Delta \tau_i$. 
If the birth rate is constant and 
selection effects are similar 
for pulsars with different spin-down ages,
then the ratio of pulsars number at different intervals of spin-down age 
reflects the ratio of true ages of pulsars in these intervals.

\begin{figure}
\includegraphics[width=\hsize]{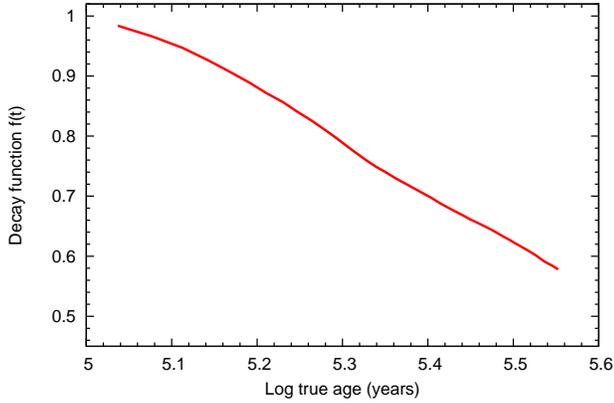}
\caption{
Magnetic field decay in normal radio pulsars from the Parkes Multibeam and
Swinburne surveys (PMSS) according to Igoshev \& Popov (2014)\nocite{ip2014}. 
The obtained result
cannot be directly extrapolated to smaller or larger ages. 
}
\label{label1}
\end{figure}

With currently available data our method is applicable to sources with ages
$\sim(1$-$5)\times 10^5$~yrs. We obtained that the field decays in
this age interval by a factor of $\sim2$. 
This can be fitted by an exponential decay with time scale $\tau\approx 4
\times 10^5$~yrs. We show the main result by Popov \& Igoshev(2014)\nocite{ip2014} in Fig.\ref{label1}. 
There (and in other plots) The magnetic field is represented as
$B(t)=B_0\times f(t)$, where $f(t)$ is the decay function which describes
field evolution.
 
Obviously, the result presented in Fig.~\ref{label1} is not in 
correspondence with most of the previous studies (see, for example, 
\cite{fgk2006} and
references therein) if it is assumed that the field
continues to diminish with the same rate at larger ages. On other hand,
moderate field decay in normal pulsars is not in contradiction with
observations (\cite{gullon}). 
So, the stage of relatively rapid decay might be terminated. 
The method is statistical in its nature, and therefore it cannot shed
light on the physical mechanism which governs the magnetic field decay at this
interval of true ages. In the new study we address this question.

\section{Hall cascade vs. Ohmic dissipation on phonons}

Basing on current understanding of NS crust properties, 
there are two effects which can cause the magnetic field
decay with a time scale $\sim 10^5$ years: the Hall effect and 
Ohmic decay 
(\cite{geppert2009}, Cumming, Arras \& Zweibel 2004\nocite{cumming2004}).
The former effect is not dissipative,
it actually works similar to a turbulent cascade and
redistributes the magnetic field energy from
large scales (for example, dipole component) to smaller scales 
(in particular, if initially there existed $l$ multipole, then the attractor
would consist of $l$ and $l+2$ poloidal components, see \cite{gc2014b} for
details). 
Intensity of the Hall effect depends on the value of initial magnetic
field. 

Magnetic field decay in young NSs can be non-monotonic. There are several
reasons for that. 
The first possibility is related to a specific property of the Hall cascade.
The rate of field decay obtained in the article by \cite{ip2014}.
is close to the expected time scale of the Hall
drift for normal magnetic fields $\sim10^{12}$~--~$10^{13}$~G. Recently it
was demonstrated
by \cite{gc2013} 
that after a relatively short stage of rapid decay
driven by the Hall cascade,  
the magnetic field reaches an equilibrium stage 
which corresponds to the so-called Hall attractor where the Hall cascade
stops. 
Later the magnetic field continues to decay mostly due to the Ohmic
dissipation. So, the field evolution is non-monotonic with several
characteristic time scales in different epochs.

The second possibility is related to existence of two regimes of Ohmic
decay. In one of them resistivity is due to scattering from phonons. This mechanism
is important till the crust of a NS is hot. In another, resistivity is
determined by impurities in the matter crust.
There is a critical temperature $T_\mathrm{U}$ when magnetic field decay due to
phonons stops. The time scale of decay on phonons can be $\sim 1$~Myr and
even smaller while a NS is hot. 
For the decay on impurities the time scale depends on the parameter of
impurities $Q$,
and it can be very long if the parameter is much smaller than
unity. 

\begin{figure}
\includegraphics[width=\hsize]{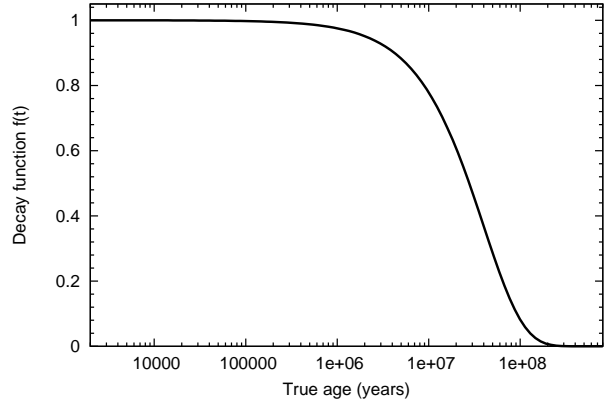}
\caption{
Magnetic field decay due to scattering from 
impurities of crystalline lattice. The impurity factor is $Q = 0.05$.
}
\label{label2}
\end{figure}

In our approach on every time step we use the formula proposed by
Aguilera, Pons \& Miralles (2008)\nocite{aguilera2008}:
\begin{equation}
B(t) = \frac{ B_0 \times 
\exp(-t/\tau_\mathrm{ohm})}{1 + \tau_\mathrm{Ohm}/\tau_\mathrm{Hall} 
[1 - \exp(-t/\tau_\mathrm{Ohm})]}.
\label{magnet_decay}
\end{equation}
In this model two distinct time scales are defined. The first one is related
to 
Ohmic decay, $\tau_\mathrm{ohm}$, and the second one --- to the Hall
cascade, $\tau_\mathrm{Hall}$. The equation also can be modified to include
some minimal value of the field, at which the decay is saturated.

Let us consider which of two effects play the more important role 
in the crust of normal radio pulsars.
The timescale of Ohmic decay is:
\begin{equation}
\tau_\mathrm{Ohm} = \frac{4\pi \sigma L^2}{c^2}.
\end{equation}
Where $\sigma$ is local electric conductivity, $L$ is the typical spatial 
scale of electric currents (can be the local pressure height, see
\cite{cumming2004}), 
and $c$ is the speed of light.

The timescale of the Hall evolution is:
\begin{equation}
\tau_\mathrm{Hall} = \frac{4\pi e n_eL^2}{cB},
\end{equation}
with $n_e$ is local electron density, $e$ is elementary charge, and $B$ is local magnetic 
field.

As both timescales have the same dependence on the spatial 
scale -- $\tau \sim L^2$, -- we can get rid of it if we consider the ratio
of two time scales:
\begin{equation}
\frac{\tau_\mathrm{Ohm}}{\tau_\mathrm{Hall}} = \frac{\sigma B}{e n_e c}.
\label{ratio}
\end{equation} 

\begin{figure}
\includegraphics[width=\hsize]{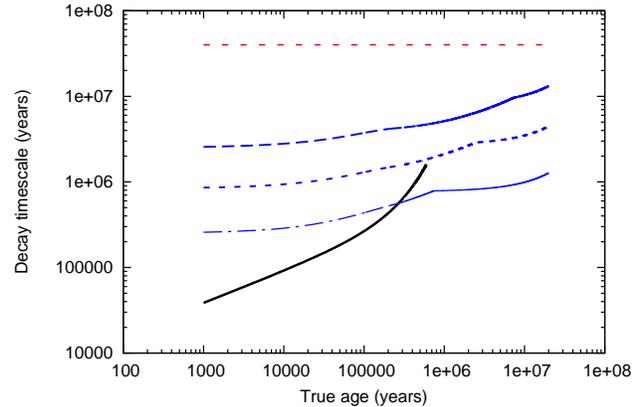}
\caption{
Typical timescales of magnetic field evolution. Thick line on the bottom
represents Ohmic decay time scale due to phonons. It is truncated when
$T_\mathrm{U}$ is reached. Dotted line on the top
corresponds to Ohmic decay due to impurities. In the middle three curves are shown for
the Hall time scales for different  initial magnetic fields. From top to
bottom: $B_0=$ $10^{12}$~G, $3 \times 10^{12}$~G, and $10^{13}$~G.   
}
\label{label4}
\end{figure}

If we use the crust properties from (\cite{cumming2004}) 
we can simplify 
Eq. (\ref{ratio}):
\begin{equation}
\frac{\tau_\mathrm{Ohm}}{\tau_\mathrm{Hall}}=2.6\times 10^{-3}\,  B_{12} \,
\rho_{13}^{1/6} \,  \left(\frac{Y_e}{0.04} \right)^{2/3} \times 
\label{ratio1}
\end{equation}
$$\times t^{0.37} \, \exp \left( 
\frac{t}{4.3\times 10^5 \, \mathrm{years}} \right). 
$$
This equation is valid till the temperature is higher than $T_\mathrm{U}$, i.e.
while  scattering from phonons is important. For fields $B \sim$ few $\times
10^{12}$~G the two time scales are equal for ages $\sim $few $\times
10^5$~yrs (see Fig.~\ref{label4}).

In our study we develop a model which includes all these types of decay
mainly following prescriptions from (\cite{cumming2004}). 
To fix parameters of the
crust we assume that the Hall time scale is equal to the scale found in
(\cite{ip2014}) for fields $\sim$few$\times 10^{12}$~G. 
To take into account
thermal evolution of NSs we use cooling curves from
(\cite{casa2011}).\footnote{We thank Peter Shternin for 
providing these data
and helpful comments.}

\begin{figure}
\includegraphics[width=\hsize]{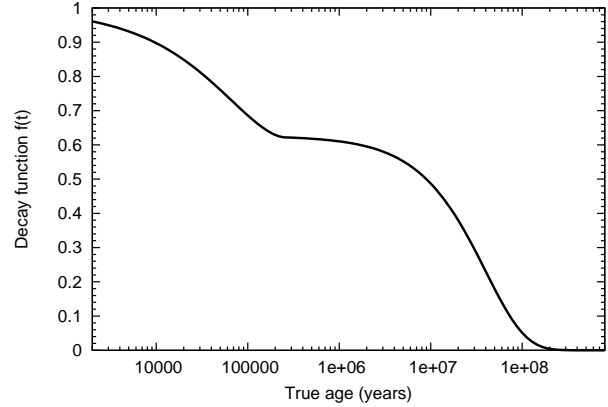}
\caption{
Ohmic magnetic field decay due to scattering from phonons and impurities.  
At the age $\sim 2\times 10^5$ years a NS cools down to the critical
temperature, $T_\mathrm{U}$,  when scattering from phonons is suppressed.
}
\label{label3}
\end{figure}

To understand the role of each types of decay we make several runs when one
or another of them are switched off. In Fig.~\ref{label2} we present the field
evolution only due to Ohmic decay via scattering from impurities. This
parameter is taken to be small: $Q=0.05$. This is a realistic assumptions
for normal field NSs, oppositely for magnetars there is a contribution from
the part of the crust in the pasta phase, where $Q$ is much larger (see
\cite{gullon} and references therein for details).
Then we include scattering from
phonons, see Fig.~\ref{label3}. 
For the chosen parameters we see rapid decay which stops at 
$\sim 2\times 10^5$~yrs. 

The time scale for the Hall effect depends on the initial field value. It can be
shorter or longer than the time scale for Ohmic decay. In
Fig.~\ref{label4} we compare the time scales for different types of decay.   
As we see, for realistic parameters the Hall effect can be as important as
Ohmic decay on phonons in the time interval of interest
$\sim(1$-$4) \times 10^5$~yrs. 

\begin{figure}
\includegraphics[width=\hsize]{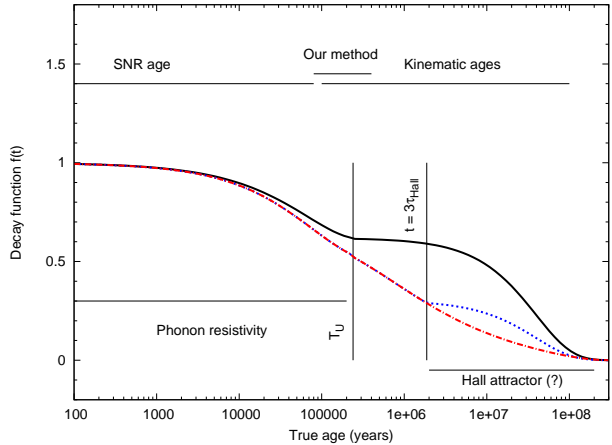}
\caption{
Magnetic field decay in three different regimes. From top to bottom: no Hall
cascade (solid), Hall cascade with the Hall attractor (dotted), 
Hall cascade without attractor (dot-dashed). For the Hall cascade
$\tau_{0\mathrm{Hall}} = 5\times 10^5$ years. The attractor stage is reached
at $3\tau_{0\mathrm{Hall}}$.
}
\label{label5}
\end{figure}

For the Hall effect we have also to consider the possibility that the Hall
attractor develops. Following \cite{gc2014}
we assume that the stage of attractor starts at $3\, \tau_{0\mathrm{Hall}}$. 
As we do not probe early evolution we neglect possible field growth at early
stages described by \cite{gc2015}.
In Fig.~\ref{label5} we present our main results.

In Fig.~\ref{label5} we present field evolution for the case without Hall
cascade, with Hall cascade but without attractor, and finally, the one with
all effects included. In addition, we plot typical ranges of time intervals
when a NS age can be estimated from the SNR nebula and from kinematic age,
and the range in which our method can be applied together with some other
 important time intervals and moments.

We see that for the chosen set of parameters in the range considered by
Igoshev \& Popov (2014)\nocite{ip2014} 
decay might be due to joint effect of the Hall cascade and
Ohmic decay on phonons. Decay is at first slowed down when temperature of NS
interiors reaches the critical value $T_\mathrm{U}$, and then (out of the
considered range) when the Hall attractor is reached. 

Note, that if close-by cooling isolated NSs (the Magnificent seven) are
descendants of magnetars (see, \cite{p2010}), then at ages
$\sim$few$\times 10^5$~yrs they can be at the stage of the Hall attractor. It would
be interesting to estimate the octupole fields of these objects.

\section{Discussion}

 In this section we have to make several comments on different assumptions
made in our preliminary study. 
In brief, the results depend significantly on parameters assumed. 
However, the main conclusion is that it is possible 
to explain the rate of field decay found in the article by
\cite{ip2014}, as well as termination of this rapid field evolution, in a
realistic physical framework. Now let us discuss it in more details.

 At first, note that $T_\mathrm{U}$ depends on several parameters, and so
decay on phonons can be terminated earlier or later. Thermal evolution
also can vary from the case we use in our calculations. This can be either
due to differences between different NSs (for example, onset of direct URCA
process), or due to some imperfections of the theoretical model behind the
results of calculations we use. As a result, if $T_\mathrm{U}$ is reached
earlier, then in the range of interest decay is mainly due to the Hall
effect. Oppositely, if $T_\mathrm{U}$  is reached later, then the Hall
effect does not dominate.

 In our study we {\it assumed} that $\tau_\mathrm{Hall}$ is compatible with
the time scale from (\cite{ip2014}). This influenced our estimates of
properties of the layer where currents are situated. If this basic
assumption is wrong, then the results might be modified significantly. We plan
to study it in the future.

 Of course, even if we can explore the whole set of parameters, we are
limited by general model assumptions, and some (micro)physics is missing. 
It would be important to probe field properties and 
evolution with observational data to put constraints on models and to guide
theoretical studies in the right direction. 

\section{Conclusions}

In this paper we presented preliminary results of the analysis of the origin
of the magnetic field decay in normal radio pulsars found by \cite{ip2014}.
We demonstrate that the decay with a time scale $\sim 4\times 10^5$~yrs and
suppression at $\sim 10^6$~yrs (or slightly earlier) can be related to two
processes. 
The first one is Ohmic decay due to phonons. It has the time scale in the 
appropriate range, and can be stopped when the NS temperature falls down to
the critical value, $T_\mathrm{U}$. The second one is related to the Hall
cascade. Fitting the depth at which electric currents are situated we can
obtain the correct time scale. The field evolution due to the Hall effect
can reach the stage of attractor at the NS age consistent with findings from
(\cite{ip2014}). However, we think that the second possibility is less
probable. May be the Hall effect also contributes to the field decay in
normal radio pulsars (see Fig.~\ref{label5}), but the main contribution,
according to our analysis, is due the Ohmic decay on phonons. A more
detailed study will be presented elsewhere (\cite{ip2015}). 

\acknowledgements
We thank Peter Shternin and his colleagues for providing the data on NS
cooling.
S.P. was supported by the RSF (grant 14-12-00146). 
S.P. is the ``Dynasty'' foundation fellow. 



\begin{thebibliography}{}
\bibitem[Aguilera et al. 2008]{aguilera}Aguilera, D.N., Pons, J.A., Miralles, J.A.: 2008, ApJ 673, L167
\bibitem[Chanmugam 1992]{chan1992}Chanmugam, G.: 1992, Ann. Rev. Astron. Astroph. 30, 143
\bibitem[Cumming et al. 2004]{cumming2004}Cumming, A., Arras, P., Zweibel, E.: 2004,
ApJ 609, 999
\bibitem[Faucher-Gigu{\`e}re \& Kaspi 2006]{fgk2006}
Faucher-Gigu{\`e}re, C.-A., Kaspi, V.~M.: 2006, ApJ 643, 332
\bibitem[Geppert 2009]{geppert2009}Geppert, U.: 2009, Astrophys. Space Sci. Lib. 357, 319
\bibitem[Gourgouliatos \& Cumming (2013)]{gc2013}Gourgouliatos, K.~N., Cumming, A.: 2013, MNRAS 434, 2480
\bibitem[Gourgouliatos \& Cumming (2014a)]{gc2014}Gourgouliatos, K.~N., Cumming, A.:
2014a, MNRAS 438, 1618
\bibitem[Gourgouliatos \& Cumming 2014b]{gc2014b}Gourgouliatos, K.~N., Cumming, A.:
2014b, Phys. Rev.
Lett. 12, 171101
\bibitem[Gourgouliatos \& Cumming (2015)]{gc2015}Gourgouliatos, K.~N., Cumming, A.: 2015, MNRAS 446, 1121 
\bibitem[Gullon et al. 2014]{gullon}Gull{\'o}n, M., Miralles, J.~A.,
Vigan{\`o}, D., Pons, J.~A.: 2014, MNRAS 443, 1891
\bibitem[Igoshev \& Popov 2014]{ip2014} Igoshev, A.P., Popov, S.B.: 2014, MNRAS 444, 1066
\bibitem[Igoshev \& Popov, in prep.]{ip2015}Igoshev, A.P., Popov, S.B.: 2015,
in preparation 
\bibitem[Mereghetti et al. 2015]{mpm2015}Mereghetti, S., Pons, J.~A., Melatos, A.: 2015, Space Sci.
Rev. 26 (in press)
\bibitem[Popov et al. 2010]{p2010} Popov, S.B., Pons, J.A., Miralles, J.A.,
Boldin, P.A.: 2010, MNRAS 401, 2675
\bibitem[Shternin et al. 2011]{casa2011}Shternin, P.~S., Yakovlev, D.~G., Heinke, C.~O.,
Ho, W.~C.~G., Patnaude, D.~J.: 2011, MNRAS 412, L108
\bibitem[Vivekanand \& Narayan 1981]{vn1981}Vivekanand, M., Narayan, R.: 1981, Journ. Astrophys. Astron.
2, 315
\bibitem[Vigan{\`o} et al. (2013)]{v2013} Vigano, D., Rea, N., Pons, J. A., Perna, R., Aguilera, D.
N., Miralles, J. A.: 2013, MNRAS 

\end{thebibliography}

\end{document}